\documentclass[12pt]{article}
\usepackage[]{epsf}
\begin{document}
\centerline{\bf\Large Thermodynamic Sampling of Molecular Conformations}
\vspace{1cm}
\centerline{\bf Andreas Kr\"amer\footnote{Email: {\tt andreas@linanthus.com}}}
\vspace{1cm}
\centerline{January 8, 2004}
\vspace{1cm}
\begin{abstract}
Torsional-space Monte Carlo simulations of flexible molecules are
usually based on the assumption that all values of dihedral angles have equal probability
in the absence of atomic interactions.
In the present paper it is shown that this assumption is not valid. Thermodynamic sampling using
dihedral angles or other internal coordinates has to account for both
the correct metric in conformational space and the
conformation-dependence of the moment of inertia tensor. Metric and moment of inertia
terms appear as conformation-dependent factors in the partition function and are obtained by
proper separation of internal and rotational degrees of freedom.
The importance of both factors is discussed for a number of short peptides
as well as for the folded and unfolded states of a protein.
It is concluded that thermodynamic Monte Carlo simulations of protein folding
that neglect these correction factors tend to underestimate the stability of the folded state.
\end{abstract}

\newpage
\section{Introduction}
Many organic molecules are able to adopt a large number of different conformations
at room temperature
which has far reaching consequences for their thermodynamic and biochemical properties.
In order to separate rotational and translational degrees of freedom, molecular conformations
are generally expressed in terms of internal coordinates where it is often sufficient to describe the
conformational state of a molecule by a set of dihedral angles since the effect of
high-frequency bond angle and length fluctuations can approximately be included in the potential energy
\cite{eff_potential}.

Dihedral angle coordinates are frequently used to sample molecular
conformations in Monte Carlo simulations of biomolecules (see for example \cite{MC_example,Favrin,
Jorgensen,Mezei}). These
thermodynamic simulations
are usually based on the assumption that the volume element appearing in the partition function
is given by
$\prod_{i=1}^Md\phi_i$ where $\phi_i$ are dihedral angles and $M$ is the number of rotatable bonds.
This means that in the absence of atomic interactions all values of dihedral
angles are sampled with equal probability. In the present paper it is shown that this underlying
assumption is generally not valid, and that the proper metric in internal coordinate space as well as
effects arising from the conformation-dependence of the moment of inertia tensor have to be taken into
account in order to accurately calculate thermodynamic quantities.

The main mathematical problem in a formulation of the statistical mechanics in the space of molecular
conformations (which we will call briefly {\em shape} space or shape manifold in the following) poses the
proper separation of {\em motions} in internal coordinates (or shape coordinates) and rotations. This is
non-trivial since it is not possible to define a body frame of reference in a unique
way as it can be done for a rigid body. In fact, one can choose an arbitrary coordinate system
for each different conformation. It is known that this freedom in the choice of the coordinate system can
be expressed in terms of a gauge potential \cite{gauge0,gauge} where local gauge transforms correspond
to independent rotations of coordinate systems, hence the gauge symmetry group is SO(3).
An interesting property of systems with internal degrees
of freedom is the possibility to generate a change in orientation
solely by the variation of shape without the application of an external torque, i.e., by moving through
a closed path in shape space. The most prominent example for this effect is the ability of a falling cat
to land on its feet starting from an upside-down position while the total angular momentum is zero.
Another example is the observation of a slow rotation of the whole system over time in
zero-angular momentum molecular dynamics simulations of proteins \cite{Karplus}.
This net rotation of the system is an example of a so-called {\em geometric phase} \cite{Berry} which
is independent of the parametrization of the closed path in shape space and also gauge-invariant.
In the case of three-dimensional molecules the practical calculation of these geometric phases is
complicated by the fact that the group of
rotations, SO(3), is non-Abelian, however, this will not be important for the following discussion.

A deeper mathematical foundation of the subject as well as generalizations can be formulated using
fiber-bundles \cite{fiber}. From a classical mechanical standpoint all
these rotation-related effects can of course also be discussed in terms of Coriolis forces.
A comprehensive recent review on the separation of internal motions and rotations in the $N$-body
problem (with emphasis on gauge fields) has recently been given by Littlejohn and Reinsch \cite{review}.

In the following section we discuss the classical canonical partition function in shape space
(its detailed derivation is given in the Appendix) and show that it
introduces two shape-dependent
correction factors which are not present in the usual Cartesian-space partition function involving
atom coordinates. The first factor transforms the volume element in shape coordinates to the true
volume element on the non-Euclidean shape manifold and involves the metric tensor. The
second term reflects the conformation-dependence of the moment of inertia tensor. Using dihedral angles
as shape coordinates the
importance of these correction terms is discussed in Section 3 for a number of small peptides,
Ace-(Ala)$_n$-Nme with $n=1,2,3$, and the pentapeptide Met-enkephalin.
In Section 4 we estimate and compare the correction terms that arise in the folded and unfolded states
of a protein.

\section{Statistical mechanics on the shape manifold}
We consider a molecule consisting of $N$ atoms with masses $m_\alpha$ for which a conformation
is uniquely described by $M$ shape coordinates $q^i$, $i=1\dots M$.
The Cartesian atom coordinates are given as functions of the shape coordinates,
\begin{equation}
\vec c_\alpha = \vec c_\alpha(q^1,\dots,q^M)
\end{equation}
with $\alpha=1,\dots,N$. The vectors $\vec c_\alpha$ are taken to be center of mass coordinates, i.e.~we assume
that for all tuples $(q^1,\dots,q^M)$ it is $\sum_\alpha m_\alpha \vec c_\alpha(q^1,\dots,q^M)=0$.
The choice of the functions $\vec c_\alpha$ is of course not unique because the atom coordinates for a
given conformation are only determined up to an overall rotation of the molecule.
In fact we could replace the functions $\vec c_\alpha$ by their arbitrarily rotated versions,
\begin{equation}
\label{gauge}
\vec c_\alpha \longrightarrow \mbox{\bf R}(q^1,\dots,q^M)\cdot \vec c_\alpha,
\end{equation}
where the rotation matrix $\mbox{\bf R}$ is an
arbitrary function of the shape coordinates. The only assumption we make is, that
$\mbox{\bf R}(q^1,\dots,q^M)$ and $\vec c_\alpha(q^1,\dots,q^M)$ are sufficiently well-behaved
w.r.t.~the existence of derivatives. Because of the shape-dependence of the atom coordinates, the
moment of inertia tensor $\mbox{\bf M}$ (Appendix Eq.~(\ref{moment_of_inertia})) is also a function of the shape
coordinates. It is convenient to write $\mbox{\bf M}$ as a dimensionless quantity,
\begin{equation}
\label{M_tilde}
\tilde{\mbox{\bf M}}=\tilde{\mbox{\bf M}}(q^1,\dots,q^M) =
\sum_\alpha \Lambda^{-2}_\alpha\left(|\vec c_\alpha|^2\mbox{\bf I}-\vec c_\alpha\otimes\vec c_\alpha\right)
\end{equation}
where
\[
\Lambda_\alpha=\left(\frac{2\pi\hbar^2\beta}{m_\alpha}\right)^{1/2}
\]
is the thermal de Broglie wavelength of atom $\alpha$ and $\beta=1/k_BT$ is the inverse temperature.
Here, $\otimes$ denotes the outer product of three-dimensional vectors. We also define the
gauge potential (see Eq.~(\ref{gauge_potential0}) in the Appendix)
\begin{equation}
\label{gauge_potential}
\vec A_i = \vec A_i(q^1,\dots,q^M)=\tilde{\bf M}^{-1}\sum_\alpha \Lambda^{-2}_\alpha\vec c_\alpha\times
\frac{\partial\vec c_\alpha}{\partial q^i},
\end{equation}
which is dimensionless by definition (assuming that the shape coordinates
$q^i$ are dimensionless), and the metric tensor $g_{ij}$ on the shape manifold
(Appendix Eq.~(\ref{metric_tensor})) that can be written in a dimensionless form as
\begin{equation}
\label{g_tilde}
\tilde g_{ij} = \tilde g_{ij}(q^1,\dots,q^M) =
\sum\limits_\alpha \Lambda^{-2}_\alpha\frac{\partial\vec c_\alpha}{\partial q_i}\cdot
\frac{\partial\vec c_\alpha}{\partial q_j} - \vec A_i\cdot\tilde{\mbox{\bf M}}\cdot \vec A_j.
\end{equation}
As discussed in the Appendix, infinitesimal distances $ds = \sqrt{g_{ij}dq^idq^j}$
correspond (up to a constant prefactor) to mass-weighted root
mean square deviations (RMSD) minimized w.r.t. rotations.
This fact can in principle be used to approximate distances in shape
space without referring to underlying coordinates.

The behavior of the gauge potential $\vec A_i$ under the gauge transform (\ref{gauge}) is given by
\begin{equation}
\label{gauge_transform}
\vec A_i \longrightarrow \mbox{\bf R}\cdot\left(\vec A_i + \vec\gamma_i\right),
\end{equation}
where $\gamma_i$ is defined by the partial derivative of the rotation $\mbox{\bf R}$ w.r.t.~the
shape coordinates,
$\partial{\mbox{\bf R}}/\partial q_i={\mbox{\bf R}}\cdot\vec\gamma_i\times \ $. With (\ref{gauge})
and (\ref{gauge_transform}) it is straightforward to see that the quantity
$\partial\vec c_\alpha/\partial q_i-\vec A_i\times\vec c_\alpha$ and hence the metric tensor
$g_{ij}$ (see Eq.~(\ref{metric_tensor}) in the Appendix) is gauge-invariant, i.e.~independent of the choice of the rotation
functions $\mbox{\bf R}(q^1,\dots,q^M)$. As noted in the
introduction it can be shown that closed paths in shape space are associated with a change
in orientation of the molecule.
According to Eq.~(\ref{omega_gauge}) in the Appendix the (gauge-dependent) infinitesimal rotation vector
$d\vec\phi_0$ associated with the variation of shape coordinates $dq^i$ is given by
$d\vec\phi_0 = -\vec A_idq^i$. The net rotation ${\bf S}$ generated by moving through a closed path in
shape space
can be calculated by accumulating these infinitesimal rotations along the path. However, because of the
non-Abelian nature of the rotation group SO(3), i.e.~the fact that rotations do not commute if their
rotation axes are different, it is only possible to express ${\bf S}$ in terms of a
path-ordered product \cite{review}. As a measurable quantity,
${\bf S}$ is of course gauge-invariant and also independent of the parametrization of the closed
path. It is possible to derive an explicit expression
for ${\bf S}$ in the case of infinitesimal small loops where a
generalized version of Stokes' theorem can be applied \cite{review}.
A consequence of the existence of orientational changes associated with closed loops in shape space is
the fact that the shape manifold defined by the metric $g_{ij}$ cannot be embedded in the
$3N$-dimensional coordinate space as a single-valued function, i.e.~the shape manifold
does not only exhibit curvature but also torsion \cite{Kleinert}.

It is shown in the Appendix that the classical canonical partition function $Z$ in shape space
reads
\begin{equation}
\label{shape_partition_function}
Z=8\pi^2\int dq^1\cdots\int dq^M (\det\tilde g_{ij})^{1/2}
                                 (\det\tilde{\bf M})^{1/2}\;\; e^{-\beta V(q^1,\dots,q^M)},
\end{equation}
where the prefactor $8\pi^2$ stems from the integration over all possible orientations of the molecule.
Note, that no
translational degrees of freedom have been taken into account in deriving $Z$ since the atom
coordinates $\vec c_\alpha$ are
defined as center of mass coordinates. However, because of the separability of center of mass
movements in the equation of motion, translational effects can simply be included by an additional
prefactor of $Z$. In the case of orientational constraints the prefactor $8\pi^2$ would
have to be modified accordingly.

The terms
$(\det\tilde g_{ij})^{1/2}$ and $(\det\tilde{\bf M})^{1/2}$ can be included
in the Boltzmann factor as effective conformation dependent energy terms $F_G$ and $F_M$,
\begin{equation}
\label{Z1}
Z=8\pi^2\int dq^1\cdots\int dq^M\;\; e^{-\beta (V+F_G+F_M)},
\end{equation}
where
\begin{equation}
\label{FG}
F_G(q^1,\dots,q^M) = -\frac{k_BT}{2}\log(\det\tilde g_{ij})
\end{equation}
and
\begin{equation}
\label{FM}
F_M(q^1,\dots,q^M) = -\frac{k_BT}{2}\log(\det\tilde{\bf M}).
\end{equation}
Below, we will calculate $F_G$ and $F_M$ for a number of molecules. In the remainder of the paper
it is assumed that the shape coordinates $q^1,\dots, q^M$ are given by the dihedral angles
$\phi_1,\dots,\phi_M$, where $M$ is the number of rotatable bonds.

The effect of the correction factors
$(\det\tilde g_{ij})^{1/2}$ and $(\det\tilde{\bf M})^{1/2}$ becomes visible in
the probability distributions $P_n(\phi)$ of the dihedral angles $\phi_n$ in the absence of
atomic interactions. If the correction factors are omitted in Eq.~(\ref{shape_partition_function})
values of the dihedral angles are uniformly distributed with $P_i(\phi)=(2\pi)^{-1}$, a fact which is
sometimes used to test microreversibility in Monte Carlo simulations with complicated concerted-rotation
move sets \cite{Jorgensen}. Here, these distributions are given by
\begin{eqnarray}
\label{angle_distribution}
P_n(\phi) &=& \frac{1}{Z}\int d\phi_1\dots\int d\phi_{n-1}\int d\phi_{n+1}\dots\int d\phi_M
\left[\det\tilde g_{ij}(\phi_1,\dots,\phi_{n-1},\phi,\phi_{n+1},\dots,\phi_M)\right]^{1/2} \times\nonumber\\
& & \times\left[\det\tilde{\bf M}(\phi_1,\dots,\phi_{n-1},\phi,\phi_{n+1},\dots,\phi_M)\right]^{1/2},
\end{eqnarray}
where
\[
Z = \int d\phi_1\dots\int d\phi_M
\left[\det\tilde g_{ij}(\phi_1,\dots,\phi_M)\right]^{1/2}
\left[\det\tilde{\bf M}(\phi_1,\dots,\phi_M)\right]^{1/2}.
\]

\section{Correction factors for small molecules}
In this section we calculate the conformation-dependent correction factors in the partition function (\ref{shape_partition_function})
and the related quantities $F_G$ and $F_M$ at $T=300$ K for the peptides Ace-(Ala)$_{1,2,3}$-Nme as well as for the
pentapeptide Met-enkephalin
with the sequence Tyr-Gly-Gly-Phe-Met. We only consider dihedral angles at fully rotatable bonds
($\omega$-angles are set to 180$^{\circ}$) that do not connect to methyl- or NH$_3^+$-groups.
Conformations of the polyalanines are parametrized by  2, 4, and 6 dihedral angles, while 17
dihedral angles have to be  taken into account for  Met-enkephalin.
Potential energies have been calculated with the Tinker software
package \cite{Tinker} using
the OPLS all-atom forcefield \cite{OPLS} (without electrostatic interaction
cut off and with $\epsilon=1$) in conjunction with a GB/SA
implicit-solvent term \cite{Still}.

Fig.~1 shows the terms $(\det{\tilde g_{ij}})^{1/2}$ and $(\det\tilde{\bf M})^{1/2}$ as functions of the
dihedral angles $\phi$ and $\psi$ for alanine dipeptide (Ace-Ala-Nme). It is seen that both terms
show relative variations of about 25\% and 30\% respectively. The corresponding variations of $F_G$ and $F_M$
are 0.17 kcal/mol and 0.21 kcal/mol. Note that only these relative variations
matter for the calculation of thermodynamic quantities. The distributions of the effective energies
$F_G$ and $F_M$ for Ace-(Ala)$_{2,3}$-Nme and Met-enkephalin
are shown in Fig.~2, 3, and 4, where each point $(F_G,F_M)$ corresponds to a low-energy molecular conformation.
For the polyalanines conformations have been obtained on an equidistant grid in dihedral
angle space with grid spacing $\Delta\phi=\pi/5$. In total $10^4$ conformations have been
sampled for Ace-(Ala)$_2$-Nme, and $10^6$ conformations for Ace-(Ala)$_3$-Nme. Of these conformations,
Fig.~2 (Fig.~3) shows those 200 (1000) with the lowest potential energies. The insets in Fig.~2 and
3 give the corresponding cumulative energy distributions.

Conformations of the larger and more flexible molecule
Met-enkephalin cannot be obtained by explicit enumeration. In this case
conformations have been generated by randomly choosing dihedral angles and subsequently applying
an annealing and energy minimization procedure that
uses short stochastic dynamics runs at decreasing temperatures followed by steepest-descend
minimization, and thereby mostly avoiding steric clashes and other high-energy situations.
This way, 2000 conformations have been generated, of which those 200 with the lowest potential energies
have been plotted in Fig.~4 along with the cumulative energy distribution shown in the inset.
It is known
that Met-enkephalin does not adopt a single conformation in aqueous solution
at room temperature \cite{Graham}, last but not least because of its biological
function as a neuro-transmitter binding to a number of different receptors.
One may therefore expect, that many conformations generated by the procedure described above can
actually be assumed by the molecule under biological conditions.

Variations $\Delta F_G$ and $\Delta F_M$ of the energies $F_G$ and $F_M$ in
Fig.~2, 3, and 4 (here defined by the difference between maximal and minimal values in the data)
strongly depend on the size of the molecule.
While for Ace-(Ala)$_2$-Nme and Ace-(Ala)$_3$-Nme these variations are small
($\Delta F_G=0.29$ kcal/mol and $\Delta F_M=0.21$ kcal/mol in the first case, and
$\Delta F_G=0.53$ kcal/mol and $\Delta F_M=0.35$ kcal/mol in the latter one),
they become more significant for Met-enkephalin, where $F_G$ and $F_M$ vary by
$\Delta F_G=2.45$ kcal/mol and $\Delta F_M=0.77$ kcal/mol according to the data plotted in Fig.~4.
Fig.~5 shows those
conformations of Met-enkephalin that correspond to the numbered data points in Fig.~4
where $F_G$ and $F_M$ and therefore $(\det{\tilde g_{ij}})^{1/2}$ and $(\det\tilde{\bf M})^{1/2}$
assume extreme values. Clearly, conformations with small factor $(\det\tilde{\bf M})^{1/2}$
exhibit a small radius of gyration and are therefore more compact than conformations with large
values of this term. The interpretation of the term $(\det{\tilde g_{ij}})^{1/2}$ is more difficult
and its values depend on the details of the molecule. However, consistent with the
pictures shown in Fig.~5 it can be said that more ``rigid'', stretched conformations lead to a smaller
factor $(\det{\tilde g_{ij}})^{1/2}$ than ``sloppier'', curved conformations, where small changes
in the dihedral angles result in larger changes in the atom coordinates. Since conformations with
large radius of gyration are usually more stretched than those with small radius of gyration
we may explain the weak negative correlation between $F_G$ and $F_M$ indicated by the straight
lines (obtained by a least squares fit) in Fig.~2, 3, and 4 (correlation coefficient
$= -0.5427$, $-0.5405$, and $-0.1511$). We do not find correlations between
the potential energy $E$ and $F_G$ or $F_M$ except for the case of
Met-enkephalin, where $E$ is negatively correlated with $F_M$ because
more compact conformations forming hydrogen bonds or
salt bridges are found to be energetically favored by several kcal/mol.

The probability distributions $P_n(\phi)$ of dihedral angles
in the absence of atomic interactions
defined in Eq.~(\ref{angle_distribution}) can be calculated by first obtaining a series of
conformations where all values of dihedral angles are sampled with equal probability. The
histograms $H_n(k)$ approximating $P_n(\phi)$, are
then given by
\[
H_n(k) = \frac{\left<\delta_k(\phi_n)(\det{\tilde g_{ij}})^{1/2}(\det\tilde{\bf M})^{1/2}\right>}
          {\left<(\det{\tilde g_{ij}})^{1/2}(\det\tilde{\bf M})^{1/2}\right>}
\]
where $\left<.\right>$ denotes an average over the series of conformations,
$k=0,\dots,K-1$ ($K$ being the number of bins), and
\[
\delta_k(\phi) = \left\{\begin{array}{lll}0 & \mbox{for} & \phi<\frac{2\pi k}{K}\\
                                          1 & \mbox{for} & \frac{2\pi k}{K} \leq \phi <\frac{2\pi(k+1)}{K}\\
                                          0 & \mbox{for} & \frac{2\pi(k+1)}{K} \leq \phi\end{array} \right. .
\]
It is $KH_n(k)\approx 2\pi P_n(\phi)$ for $\phi\in[\frac{2\pi k}{K},\frac{2\pi(k+1)}{K})$. Fig.~6
shows the so-obtained distributions for all 17 dihedral angles of Met-enkephalin using $10^6$
randomly sampled conformations. It is seen that deviations from the uniform distribution
$P_n(\phi)=(2\pi)^{-1}$ of up to 40\% occur for some of the dihedral angles (most prominently
for $\phi_3$, $\phi_{11}$, and $\phi_{16}$, where the indices refer to the bond indices
shown in Fig.~5). This again demonstrates that
the correction factors $(\det{\tilde g_{ij}})^{1/2}$ and $(\det\tilde{\bf M})^{1/2}$ are significant for
Met-enkephalin.

\section{Protein folding}
In this section we discuss the terms $F_G$ and $F_M$ in the context of folded and unfolded states
of a protein with radii of gyration $R^{\mbox{\tiny folded}}$ and $R^{\mbox{\tiny unfolded}}$. At first,
an estimate of $F_G$ is given considering only backbone dihedral angles ($\phi$ and $\psi$-angles)
with the assumption that the contributions
of side chain dihedrals to the metric tensor $g_{ij}$ are roughly the same for folded and
unfolded states.
In order to calculate the determinant of $g_{ij}$ we have to estimate its eigenvalues $\lambda$.
One may argue, that the corresponding eigenvectors are localized on the peptide chain and span about
seven adjacent torsion angles, having in mind that six is the number of angles needed to solve the
so-called rebridging problem for polymer chains \cite{bridging}. This means, that there is one
free parameter in the concerted motion of seven adjacent $\phi/\psi$-angles leaving the rest of the
protein unchanged. We therefore assume that an eigenvector of $g_{ij}$ corresponds to a loop of
about seven residues with length $\ell$ and with endpoints $A$ and $B$. Let $r$ be the average distance
of $A$ and $B$, and $\xi$ the length scale of the average lateral fluctuations of such a loop
w.r.t.~the axis $\vec{AB}$. Since this is a short loop (w.r.t. to the persistence length scale) we estimate
the scaling behavior of $\xi$ to be $\xi\sim\sqrt{\ell^2-r^2}$. The corresponding eigenvalue
of $g_{ij}$ should scale as $\lambda\sim m\xi^2$, where $m$ is the mass of the loop, and therefore
$\det g_{ij}\sim(m\xi^2)^{2N}$, where $N$ is the number of residues.
Assuming $r\ll\ell$ and taking $r$ to be proportional to the radius of gyration we obtain
\[
\Delta F_G = F_G^{\mbox{\tiny unfolded}} - F_G^{\mbox{\tiny folded}}\approx
Nk_BT\left(\frac{r^{\mbox{\tiny folded}}}{\ell}\right)^2\left[\left(
\frac{R^{\mbox{\tiny unfolded}}}{R^{\mbox{\tiny folded}}}\right)^2-1\right].
\]
Using a value of 1.6 for the ratio of the radii of gyration in the unfolded
and folded states from \cite{radius_of_gyration},
and very roughly estimating $r^{\mbox{\tiny folded}}/\ell\approx 0.4$ from the analysis of C$_\alpha$
distances of protein structures,
we find $\Delta F_G\approx 7.5$ kcal/mol at $T=300$ K for a protein with $N=50$ residues.
One may therefore conclude that energy corrections due to the metric of the dihedral angle shape
manifold are a significant energetic contribution and
should be taken into account in folding simulations using Monte Carlo moves based on dihedral angles.

We now turn to the term $\det\mbox{\bf M}$. Since this term arises form the thermal equilibration
of angular momentum its contribution is for instance not included in molecular dynamics
simulations that enforce $L=0$ for the protein. In order to estimate $F_M$ we have to consider inertial
effects of the solvent as well. While these effects should be negligible for small molecules it is reasonable to assume
that
in the unfolded state of a globular protein many water molecules are effectively trapped and therefore
rigidly coupled. For our estimate we therefore consider two limiting cases: complete viscous coupling
of the solvent (i.e. neglecting inertial effects of the solvent at all) and complete rigid coupling,
(i.e.~all solvent molecules in the sphere defined by the radius of gyration $R$ are rigidly coupled to
the protein). In the first case the components $M$ of the diagonalized moment of inertia tensor
$\mbox{\bf M}$ scale as
$M\sim mR^2$ where $m=const.$ is the mass of the protein, while in the latter case the mass $m$ itself
scales as
$m\sim R^3$, so that $M\sim R^5$ assuming a uniform mass density of protein and solvent.
With $\det\mbox{\bf M}\sim(mR^2)^3$ we derive
\[
\Delta F_M = F_M^{\mbox{\tiny unfolded}} - F_M^{\mbox{\tiny folded}}\approx
-k_BT\kappa\log\left(\frac{R^{\mbox{\tiny unfolded}}}{R^{\mbox{\tiny folded}}}\right)
\]
where $\kappa=3$ in the case of viscously coupled solvent and $\kappa=15/2$ for rigid coupling.
At $T=300$ K we
get $\Delta F_M\approx -0.85$ kcal/mol in the first case and $\Delta F_M\approx -2.1$ kcal/mol in the
second.

Note, that $\Delta F_G$ and $\Delta F_M$ have different signs: Unfolded or stretched conformations
have a smaller value of $\det g_{ij}$ because the loops are stiffer and therefore dihedral angle
variations result in smaller variations in Cartesian space, while the moment of inertia is the larger the
larger the characteristic length scale of the molecule is. This confirms the trend that had already
been observed for small molecules discussed in the previous section. It is seen that $|\Delta F_G|$
is by about a factor of 5 larger than $|\Delta F_M|$ for a protein with 50 residues.
While $\Delta F_G$ depends linearly on the chain length, $\Delta F_M$ is independent of the
size of the protein.

\newpage
\section{Discussion and conclusion}
It has been shown in the present paper that statistical mechanical sampling of molecular conformations
has to account for the
correct metric $g_{ij}$ in conformational space as well as the conformation-dependence of the
moment of inertia tensor $\bf M$, both of which can be expressed in terms of
effective conformation-dependent energy contributions, $F_G$ and $F_M$.
Using dihedral angles as internal coordinates, the distribution
of these energy contributions for low-energy conformations has been calculated numerically
for a number of short peptides.
While their influence is small for molecules with few rotatable bonds, we
find variations of $F_G$ and $F_M$ of about 2.45 kcal/mol (for $F_G$) and 0.77 kcal/mol (for $F_M$) for the pentapeptide
Met-enkephalin. A rough estimate of both terms in the folded and unfolded states of a protein with
50 residues leads to the significant energy difference of 7.5 kcal/mol for $F_G$ and between
-0.85 kcal/mol and -2.1 kcal/mol for $F_M$. This shows that both correction terms should be taken into
account when dihedral coordinates are used in thermodynamic Monte Carlo simulations
of protein folding in order to accurately calculate thermodynamic quantities.
Since $F_G+F_M$ is larger in the unfolded state of a protein, Monte Carlo
simulations that omit these corrections lead to free energy differences
between unfolded and folded states that are too small. These simulations therefore
underestimate the stability of the folded state.
The efficient implementation of metric and moment of inertia-related correction terms within a
Monte Carlo algorithm will be subject of a future publication.

A related problem, where the consideration of the proper metric in conformational space is important,
is the estimate of
thermodynamic quantities from a given, finite ensemble of molecular conformations.
Such ensembles can be generated for
molecules with not too many rotatable
bonds and allow for the calculation of free energies and conformational entropies which are otherwise
difficult to
access in importance sampling-based methods.
An example is the calculation of protein side chain free energies \cite{Mayo}
from rotamer libraries \cite{Dunbrack}. In principle it is possible to estimate
thermodynamic properties such as entropies from any given ensemble of conformations without
referring to underlying internal coordinates. Distances
according to the metric $g_{ij}$ can be approximated by the mass-weighted RMSD between two
conformations minimized w.r.t.~rotations and translations. However, the calculation of the partition function would then require
a triangulation procedure in order to calculate volume elements in conformational space.

\section*{Appendix}
In the following we will derive the classical canonical partition function $Z$ in shape space.
At first, an expression for the Hamiltonian $\cal H$ is obtained which separates shape space
and rotational contributions. A more detailed derivation of $\cal H$ and a discussion of the related
theory can be found in the review article \cite{review}.
In order to keep equations as simple as possible three different notations are used to
distinguish between vectors in three-dimensional Cartesian space, sums over atom positions
$\vec c_\alpha$, $\alpha=1\dots N$ and sums over shape coordinates $q^i$, $i=1\dots M$. Three-dimensional
vectors are given in a vector notation with dots ($\cdot$) and crosses ($\times$) for scalar and vector
products. The Einstein sum convention is employed for summation over latin indices involving the shape
coordinates $q^i$. Sums over atom positions (greek indices) are written explicitly. We also define
a $3N$-dimensional vector (denoted in bold face), $\mbox{\bf c} = (\vec c_1,\dots,\vec c_N)$,
that contains the (three-dimensional) atom positions as components.
Vector operations, $\cdot$ and $\times$, acting on 3$N$-dimensional vectors are meant to act on each
vector component independently. Finally, $<|>$ is a scalar product in the $3N$-dimensional vector space
defined by
\begin{equation}
\label{scalar_product}
<\mbox{\bf u}|\mbox{\bf v}>=\sum_\alpha m_\alpha\vec u_\alpha\cdot\vec v_\alpha,
\end{equation}
where the associated norm is $\|\mbox{\bf u}\|^2=<\mbox{\bf u}|\mbox{\bf u}>$.

Let a molecular conformation be given by a vector ${\bf c}$ and
consider a (kinematic) variation of shape coordinates $\{dq^i\}$ and the resulting variation
$d\mbox{\bf c}$ in Cartesian space,
\begin{equation}
d\mbox{\bf c}=\left(\frac{\partial\vec c_1}{\partial q_i}d q^i,\dots,
\frac{\partial\vec c_N}{\partial q_i}d q^i   \right).
\end{equation}
The goal is to separate $d\mbox{\bf c}$ into a pure shape variation part
$\mbox{\bf c}_\parallel$ and a pure rotational part $\mbox{\bf c}_\perp$,
\begin{equation}
d\mbox{\bf c}=d\mbox{\bf c}_\parallel+d\mbox{\bf c}_\perp,
\end{equation}
 such that both parts are orthogonal to each other
\begin{equation}
\label{orthogonality}
<d\mbox{\bf c}_\parallel|d\mbox{\bf c}_\perp>=0.
\end{equation}

This can be achieved by defining an infinitesimal rotation $\mbox{\bf R}=\mbox{\bf I}+d\vec\phi\times$
that acts on $\mbox{\bf c}$ where $d\vec\phi$ is an angular rotation vector. Then,
\[
\mbox{\bf R}\mbox{\bf c}-\mbox{\bf c} = d\vec\phi\times\mbox{\bf c}
\]
defines a three-dimensional linear subspace $V$ which is parametrized by $d\vec\phi$, and
$d\mbox{\bf c}_\perp$ is the orthogonal projection
of $d\mbox{\bf c}$ on $V$. The vector
$d\mbox{\bf c}_\perp$ can be calculated by minimizing the infinitesimal distance
\begin{equation}
\label{minimum}
D(d\vec\phi) = \|d\mbox{\bf c} + d\vec\phi\times\mbox{\bf c}\|,
\end{equation}
where the minimum $D_0$ is defined by
\begin{equation}
\label{minimum1}
D_0 = D(d\vec\phi_0) = \min_{d\vec\phi}\left\{D(d\vec\phi)\right\}.
\end{equation}
This leads to the result
\begin{equation}
d\vec\phi_0=\mbox{\bf M}^{-1}\cdot\sum_\alpha m_\alpha d\vec c_\alpha\times\vec c_\alpha
\end{equation}
where
\begin{equation}
\label{moment_of_inertia}
\mbox{\bf M}=\sum_\alpha m_\alpha\left(|\vec c_\alpha|^2
\mbox{\bf I}-\vec c_\alpha\otimes\vec c_\alpha\right)
\end{equation}
is the moment of inertia tensor, and we have
\begin{equation}
\label{c_perp}
d\mbox{\bf c}_\perp=-d\vec\phi_0\times\mbox{\bf c}
\end{equation}
and
\begin{equation}
\label{c_parallel}
d\mbox{\bf c}_\parallel=d\mbox{\bf c}+d\vec\phi_0\times\mbox{\bf c}.
\end{equation}
The rotation vector $d\vec\phi_0$ can be expressed in terms of a gauge potential $\vec A_i$,
\begin{equation}
\label{omega_gauge}
d\vec\phi_0 = -\vec A_idq^i,
\end{equation}
where $\vec A_i$ is defined as
\begin{equation}
\label{gauge_potential0}
\vec A_i = \mbox{\bf M}^{-1}\sum_\alpha m_\alpha\vec c_\alpha\times
\frac{\partial\vec c_\alpha}{\partial q^i}.
\end{equation}
Note, that the variation $d\mbox{\bf c}_\parallel$ is
independent of the freedom in the choice of the coordinate functions $\vec c_\alpha(q^1,\dots,q^M)$
described in Eq.~(\ref{gauge}) because it has been obtained by minimizing the distance
$D$, while $d\mbox{\bf c}_\perp$ is non-unique and depends on this choice. According to
Eq.~(\ref{minimum}) and (\ref{minimum1}) the infinitesimal distance $D_0$ corresponds to the
{\em mass-weighted RMSD}
between ${\bf c}+d{\bf c}$ and ${\bf c}$ minimized w.r.t.~rotations.

Let us now use a body frame of reference and assume that the Cartesian coordinates of
the system are momentarily described by the vector $\mbox{\bf c}$. Consider a
variation of Cartesian coordinates $d\mbox{\bf r}$  which is the sum of
a shape contribution $d\mbox{\bf c}$ and an external infinitesimal rotation given by
$d\vec\phi\times\mbox{\bf c}$,
\[
d\mbox{\bf r} = d\mbox{\bf c} + d\vec\phi\times\mbox{\bf c}.
\]
Using Eqs.~(\ref{c_perp}) and (\ref{c_parallel}) we can write
\begin{equation}
d\mbox{\bf r}=d\mbox{\bf c}_\parallel+(d\vec\phi-d\vec\phi_0)\times\mbox{\bf c}.
\end{equation}
Because the rotational part of $d\mbox{\bf r}$, $(d\vec\phi-d\vec\phi_0)\times\mbox{\bf c}$, is an
element of the linear space $V$, it
is also orthogonal to $d\mbox{\bf c}_\parallel$. This can be independently verified
using the definition of the scalar product (\ref{scalar_product}).

The kinetic energy of the system is given by
\[
T=\frac{1}{2}\sum\limits_\alpha m_\alpha\left(\frac{d\vec r_\alpha}{dt}\right)^2
\]
which can be expressed in terms of the scalar product (\ref{scalar_product}) as,
\[
T=\frac{1}{2}\left<\frac{d\mbox{\bf r}}{dt}\left|\frac{d\mbox{\bf r}}{dt}\right.\right>.
\]
The orthogonality between the shape and rotational part of $d\mbox{\bf r}$ described above
now allows us to separate both contributions in the expression for $T$,
\[
T = \frac{1}{2}\sum\limits_\alpha m_\alpha\left\{
\left(\frac{\partial\vec c_\alpha}{\partial q_i}-\vec A_i\times\vec c_\alpha\right)\cdot
\left(\frac{\partial\vec c_\alpha}{\partial q_j}-\vec A_j\times\vec c_\alpha\right)\dot q^i\dot q^j
+\left(\vec\omega\times\vec c_\alpha + \vec A_i\times\vec c_\alpha\dot q^i\right)^2\right\},
\]
where the angular velocity vector is defined by $\vec\omega=d\vec\phi/dt$ and we inserted
the expressions~(\ref{c_perp}), (\ref{c_parallel}), and (\ref{omega_gauge}).
With the definition of the metric tensor
\begin{equation}
\label{metric_tensor}
g_{ij} =
\sum\limits_\alpha m_\alpha
\left(\frac{\partial\vec c_\alpha}{\partial q_i}-\vec A_i\times\vec c_\alpha\right)\cdot
\left(\frac{\partial\vec c_\alpha}{\partial q_j}-\vec A_j\times\vec c_\alpha\right) =
\sum\limits_\alpha m_\alpha\frac{\partial\vec c_\alpha}{\partial q_i}\cdot
\frac{\partial\vec c_\alpha}{\partial q_j} - \vec A_i\cdot\mbox{\bf M}\cdot \vec A_j
\end{equation}
the kinetic energy $T$ can finally be written as
\begin{equation}
\label{kinetic_energy}
T = \frac{1}{2}g_{ij}\dot q^i\dot q^j + \frac{1}{2}(\vec\omega+\vec A_i\dot q^i)\cdot \mbox{\bf M} \cdot
(\vec\omega+\vec A_j\dot q^j).
\end{equation}
We now assume that the system is only subject to forces between the atoms, which means that
the potential energy $V=V(q^1,\dots,q^M)$ is a function of the shape coordinates alone and the
Lagrangian can be written as
\begin{equation}
\label{Lagrangian}
{\cal L} = T - V(q^1,\dots,q^M).
\end{equation}
From Eq.~(\ref{kinetic_energy}) the angular momentum $\vec L$ is obtained by
\begin{equation}
\label{angular_momentum}
\vec L=\frac{\partial T}{\partial\vec\omega} = \mbox{\bf M}\cdot(\vec\omega+\vec A_i\dot q^i),
\end{equation}
and the generalized momenta $p_i$ associated with the shape
coordinates $q^i$ are
\begin{equation}
\label{generalized_momenta}
p_i=\frac{\partial T}{\partial \dot q^i} = g_{ij}\dot q^j + \vec L\cdot A_i.
\end{equation}
In the case of zero angular momentum, i.e.~the absence of any external torque, the rotation generated
by an internal motion $\dot q^i$ is therefore given by $\vec\omega=-\vec A_i\dot q^i$. Comparing this
with Eq.~(\ref{omega_gauge}) shows
that the shape manifold itself is actually defined by the condition $\vec L=0$ \cite{review}.

Now, the Hamiltonian ${\cal H} = T+V$ can be obtained by
solving (\ref{angular_momentum}) and (\ref{generalized_momenta}) for the angular and shape coordinate
velocities and inserting them into the
kinetic energy (\ref{kinetic_energy}),
\begin{equation}
\label{Hamiltonian}
H = \frac{1}{2}g^{ij}(p_i-\vec L\cdot\vec A_i)(p_j-\vec L\cdot\vec A_j)+\frac{1}{2}
\vec L\cdot \mbox{\bf M}^{-1}\cdot\vec L + V(q^1,\dots,q^M).
\end{equation}
Here, $g^{ij}$ is the inverse metric tensor, and $q^i$ and $p_i$
are canonically conjugated variables. It is not
possible to define coordinates which are canonically conjugated to the angular momentum $\vec L$
\cite{review}, because
the angular velocity $\vec\omega$ cannot be written as a simple time derivative of angular coordinates.
Nevertheless,
in order to derive the partition function $Z$ from $\cal H$, it is necessary to describe the rotational
part of the Hamiltonian $\cal H$ with canonical variables as well.
This can be achieved by using Euler angles $\theta$, $\varphi$, $\psi$ and their canonically conjugated
momenta $p_\theta$, $p_\varphi$, $p_\psi$. Here, it is convenient to use the principle
axes of the moment of inertia tensor ${\bf M}$ as the basis of the
underlying coordinate system (where ${\bf M}$ is diagonal). With
\[
\vec\omega = \left(\begin{array}{c}\dot\theta\cos\psi+\dot\psi\sin\psi\sin\theta\\
\dot\theta\sin\psi-\dot\psi\cos\psi\sin\theta\\
\dot\varphi+\dot\psi\cos\theta\end{array}\right)
\]
we have
\[
\vec L = \frac{1}{\sin\theta}\left(\begin{array}{c}
\cos\psi\sin\theta \;p_\theta + \sin\psi(p_\varphi-\cos\theta \;p_\psi)\\
\sin\psi\sin\theta \;p_\theta - \cos\psi(p_\varphi-\cos\theta \;p_\psi)\\
\sin\theta\;p_\psi\end{array}\right),
\]
and the rotational part of kinetic energy, $\frac{1}{2}\vec L\cdot \mbox{\bf M}^{-1}\cdot\vec L$, becomes
\[
T = \frac{1}{2}\left\{\frac{[(p_\varphi-p_\psi\cos\theta)\sin\psi+p_\theta\sin\theta\cos\psi]^2}
{M_1\sin^2\theta}
+\frac{[(p_\varphi-p_\psi\cos\theta)\cos\psi-p_\theta\sin\theta\sin\psi]^2}
{M_2\sin^2\theta}+\frac{p_\psi^2}{M_3}\right\},
\]
where $M_1$, $M_2$, and $M_3$ are the diagonal components of the moment of inertia tensor.
The classical canonical partition function $Z$ of the system is then given by (no sum convention!)
\begin{equation}
\label{partition_function}
Z = \int\cdots\int\left(\prod\limits_{i=1}^M\frac{dp_idq^i}{2\pi\hbar}\right)
\frac{dp_\theta dp_\varphi dp_\psi d\theta d\varphi d\psi}{(2\pi\hbar)^3}
e^{-\beta H(q^1,\dots,q^M,p_1,\dots,p_M,\theta,\varphi,\psi,p_\theta,p_\varphi,p_\psi)},
\end{equation}
where $\beta$ is the inverse temperature.
The factor $\left(\frac{1}{2\pi\hbar}\right)^{M+3}$ comes from the standard ``coarse-graining'' procedure
in phase space which is a way to derive the correct quantum-mechanical
prefactor of classical
partition functions \cite{textbook_statmec}. In this procedure, as a consequence of the Heisenberg
uncertainty relation, the phase space is devided into cells with volume $\Delta p\Delta q=2\pi\hbar$, where $p$ and $q$ are arbitrary pairs of conjugated generalized canonical coordinates and momenta.

In order to obtain the partition function in the shape coordinates $q^i$ alone we
first integrate (\ref{partition_function}) over the generalized momenta $p_i$,
followed by an integration over the Euler momenta $p_\theta$, $p_\varphi$, $p_\psi$.
These integrals a purely Gaussian and can readily be performed analytically. Finally, the
integration over all orientations of the system, i.e.~the Euler angles themselves, is trivial and
results in a numerical factor $8\pi^2$. This leads to the final result
\[
Z = 8\pi^2\left(\frac{1}{2\pi\hbar^2\beta}\right)^{\frac{M+3}{2}}
\int dq^1\cdots\int dq^M\left(\det g_{ij}\right)^{1/2}\left(
\det\mbox{\bf M}\right)^{1/2}\;\; e^{-\beta V(q^1,\dots,q^M)}.
\]
It is seen, that the factor $(\det g_{ij})^{1/2}$ which depends
on the shape coordinates $q^i$ results from the equilibration of the  momenta in
shape space, while the factor $(\det\mbox{\bf M})^{1/2}$ originates from the equilibration
of angular momentum. Upon absorbing the factor $\left(\frac{1}{2\pi\hbar^2\beta}\right)^{\frac{M+3}{2}}$
in the determinants we arrive at the expression for $Z$ given in Eq.~(\ref{shape_partition_function}).

\newpage
\centerline{\epsfxsize 5in \epsfbox{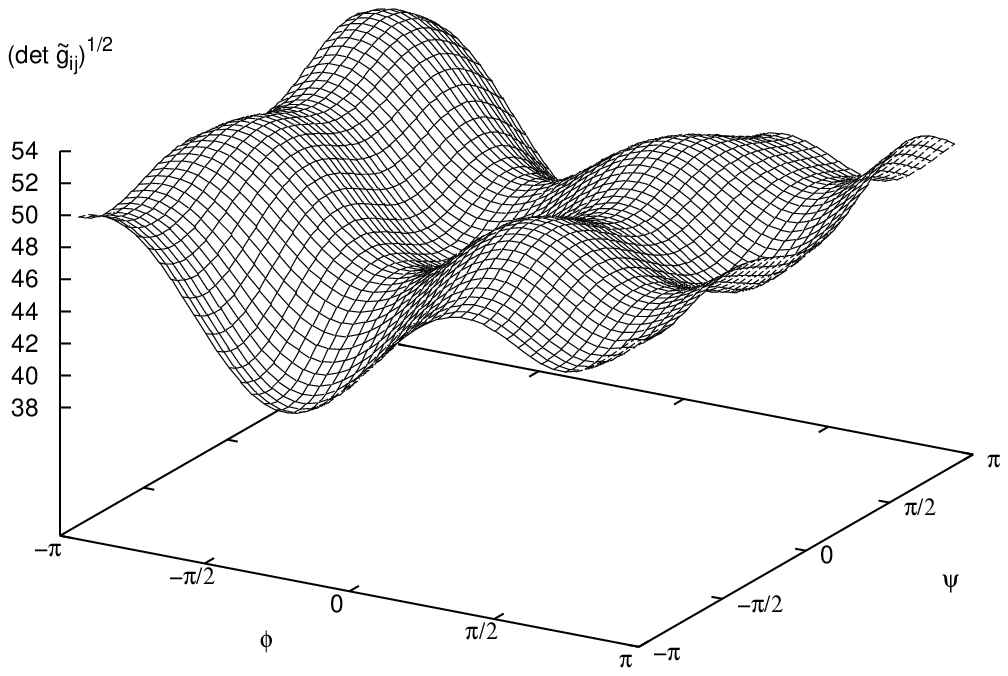}}
\centerline{\epsfxsize 5in \epsfbox{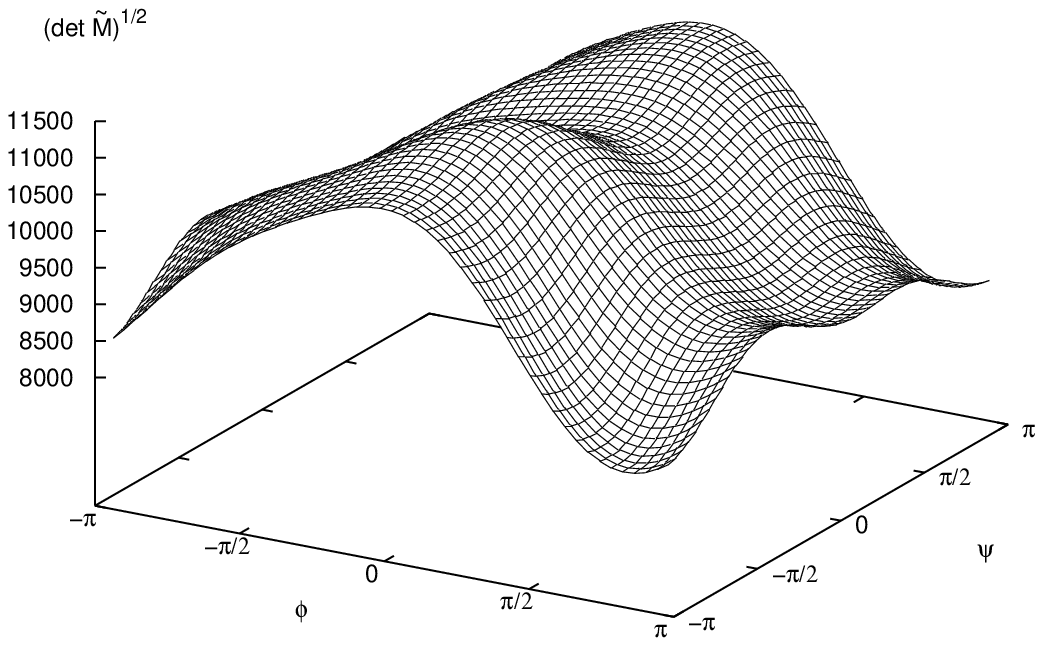}}

\vspace{1cm}
\begin{center}
\parbox[t]{6in}{\footnotesize
{\em Fig.~1.} The terms $(\det\tilde g_{ij})^{1/2}$ (top) and $(\det{\tilde{\bf M}})^{1/2}$ (bottom)
as function of the dihedral angles $\phi$ and $\psi$ for alanine dipeptide (Ace-Ala-Nme).
}
\end{center}

\newpage
\centerline{\epsfxsize 6in \epsfbox{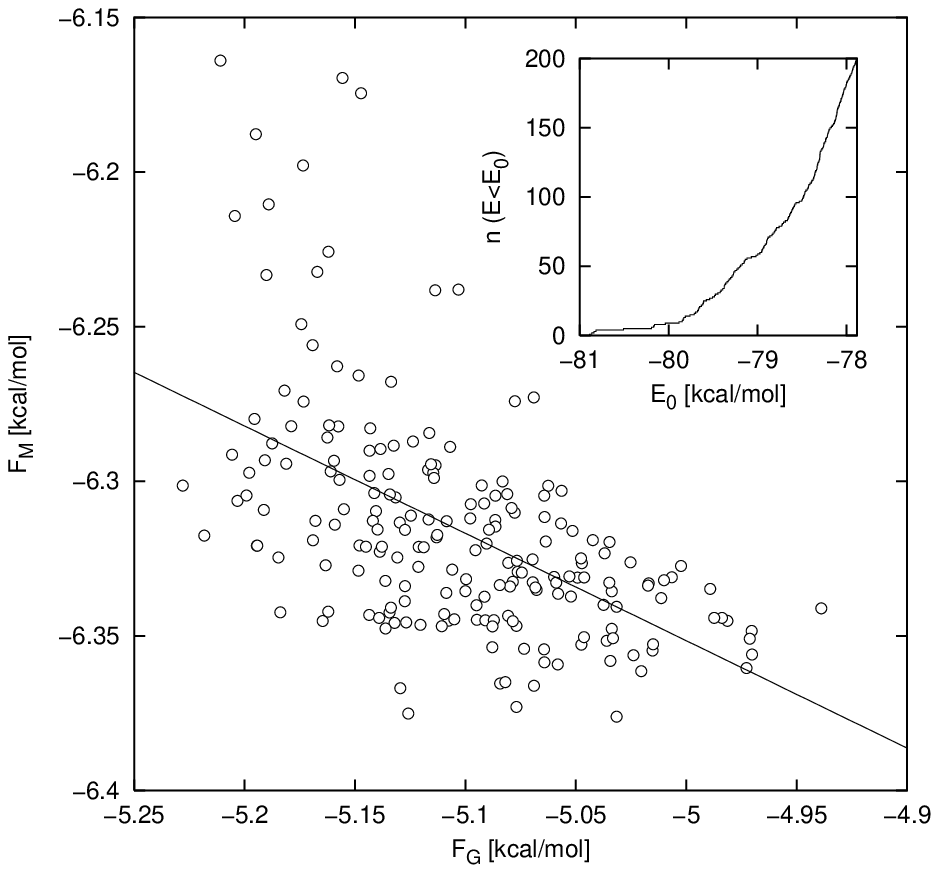}}

\vspace{1cm}
\begin{center}
\parbox[t]{6in}{\footnotesize
{\em Fig.~2.} Effective energies $F_G$ and $F_M$ for low-energy conformations of
Ace-(Ala)$_2$-Nme. The inset shows the cumulative distribution of potential energies of the sampled
conformations.
}
\end{center}

\newpage
\centerline{\epsfxsize 6in \epsfbox{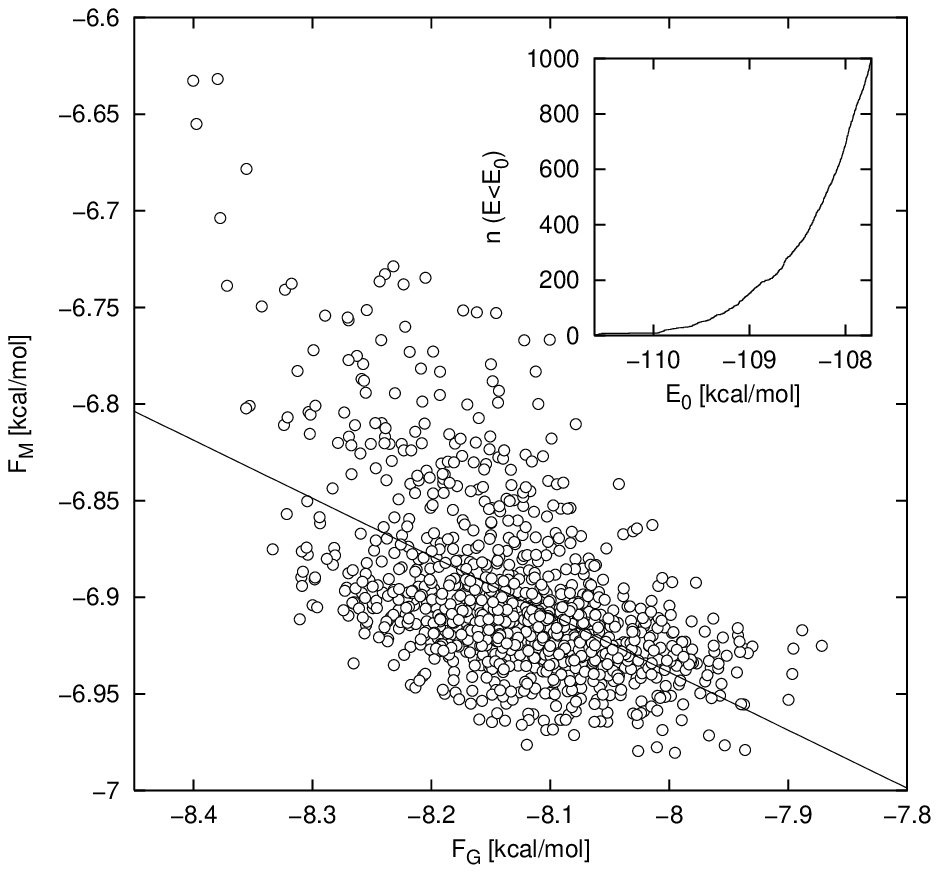}}

\begin{center}
\parbox[t]{6in}{\footnotesize
{\em Fig.~3.} Effective energies $F_G$ and $F_M$ for low-energy conformations of
Ace-(Ala)$_3$-Nme. The inset shows the cumulative distribution of potential energies of the sampled
conformations.
}
\end{center}

\newpage
\centerline{\epsfxsize 6in \epsfbox{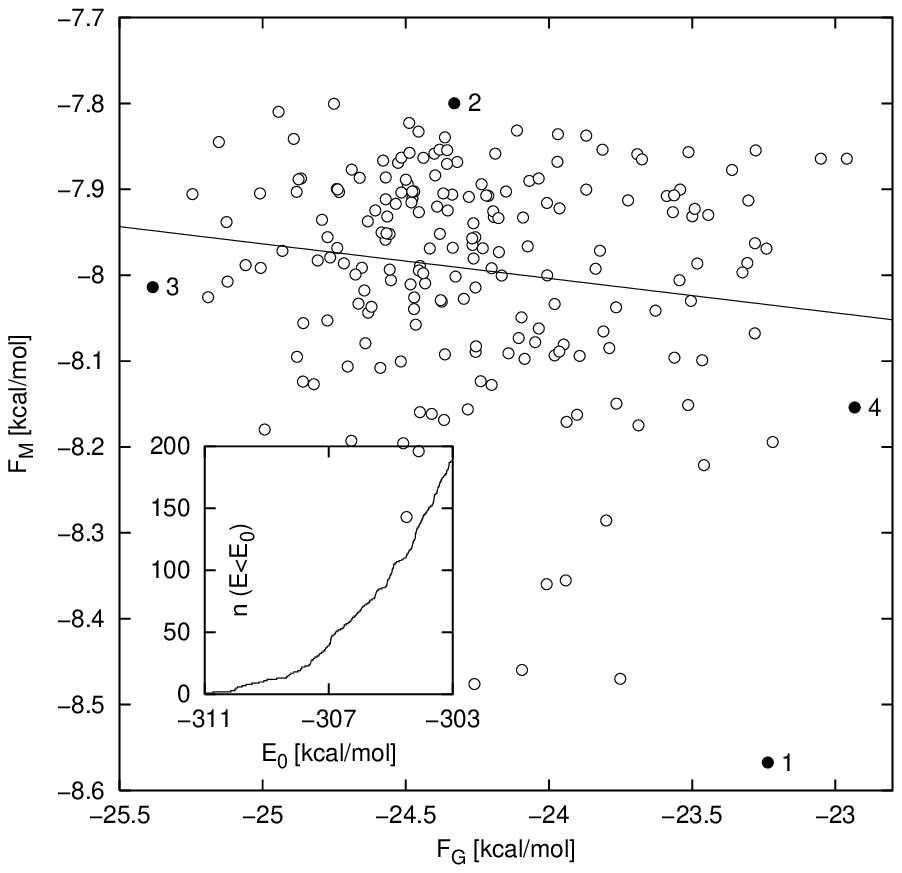}}

\vspace{1cm}
\begin{center}
\parbox[t]{6in}{\footnotesize
{\em Fig.~4.} Effective energies $F_G$ and $F_M$ for low-energy conformations of
Met-enkephalin. The inset shows the cumulative distribution of potential energies of the sampled
conformations.
}
\end{center}

\newpage
\centerline{\epsfxsize 6in \epsfbox{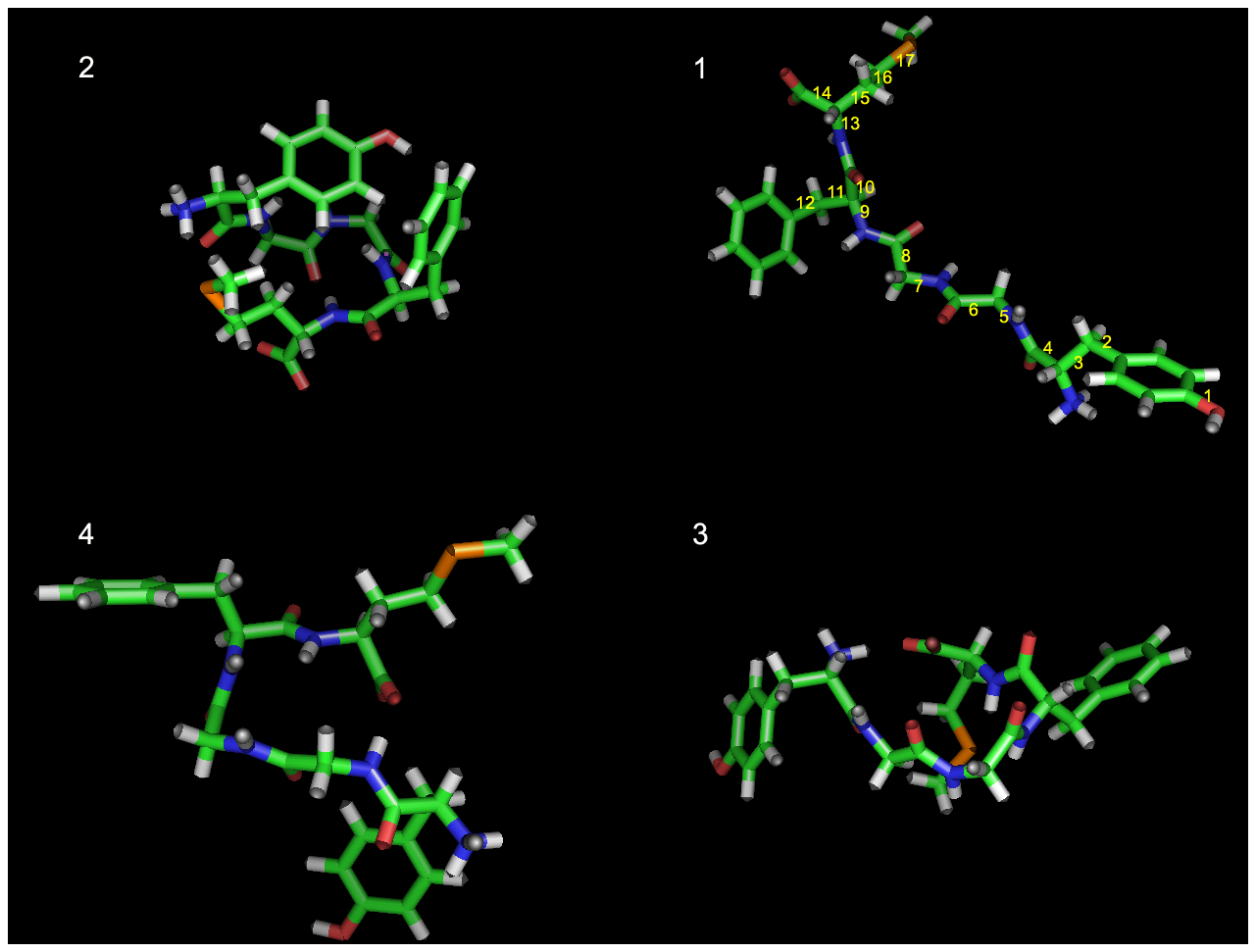}}

\vspace{1cm}
\begin{center}
\parbox[t]{6in}{\footnotesize
{\em Fig.~5.} Conformations of Met-enkephalin corresponding to the numbered data points in Fig.~4
for which either $F_G$ or $F_M$ take maximal or minimal values.
}
\end{center}

\newpage
\centerline{\epsfxsize 3in \epsfbox{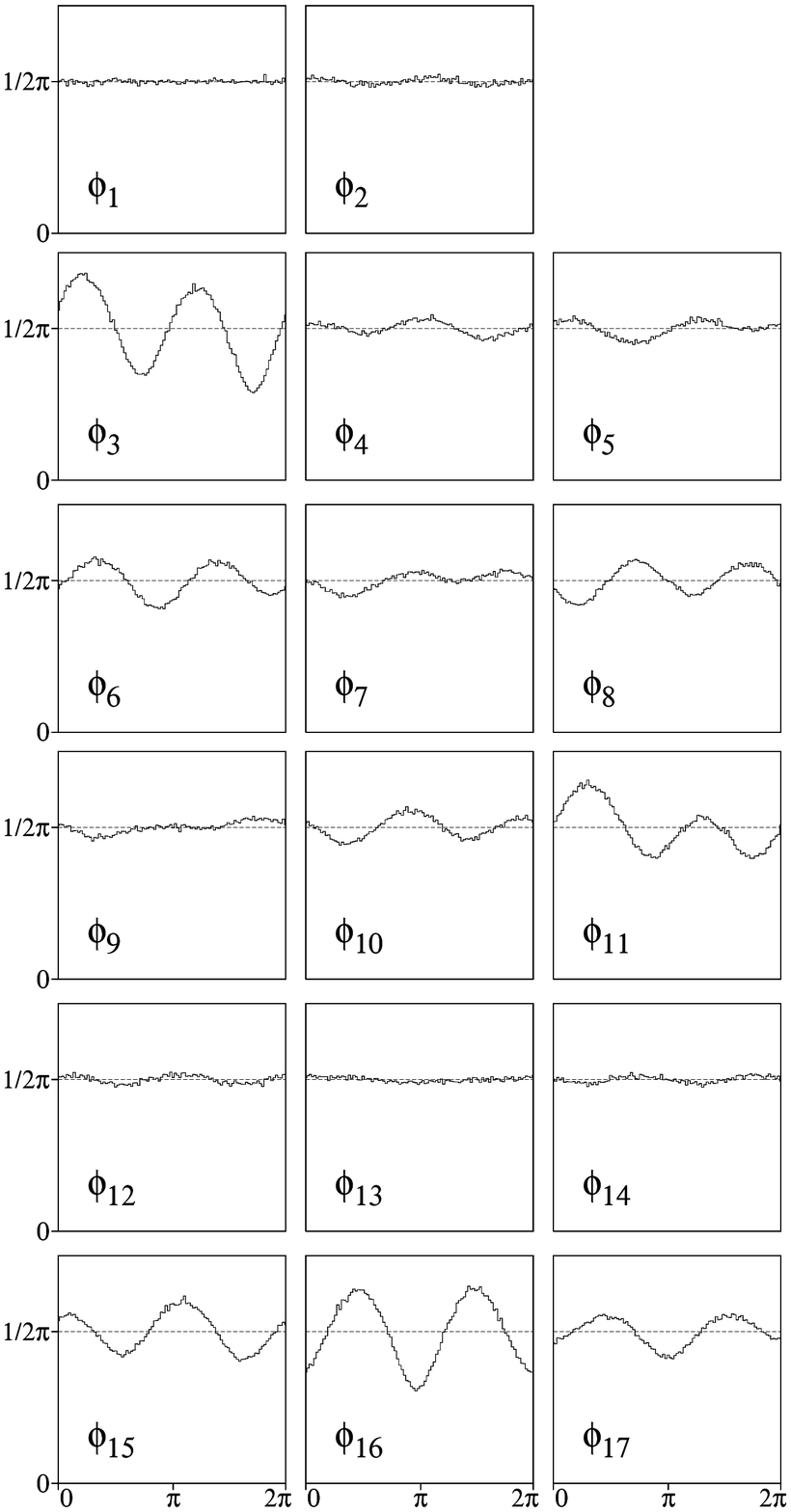}}

\vspace{1cm}
\begin{center}
\parbox[t]{6in}{\footnotesize
{\em Fig.~6.} Probability distributions of dihedral angles $\phi_i$ for Met-enkephalin
in the absence of atomic interactions. The indices $i$ correspond to the bond indices given
in Fig.~5.
}
\end{center}

\end{document}